\documentclass[aps,article,a4paper,10pt,notitlepage,footinbib,superscriptaddress,longbibliography]{revtex4-1}
\usepackage[english]{babel}
\usepackage{lmodern}
\usepackage[applemac]{inputenc}
\usepackage{amssymb,amsmath,amsfonts}
\usepackage{textcomp}
\usepackage{braket}
\usepackage{ulem}
\usepackage{soul}
\usepackage{graphicx}
\usepackage{dcolumn}
\usepackage{bm}
\usepackage{xcolor}
\usepackage{soul}
\usepackage{hyperref}
\usepackage{comment}
\usepackage{float}
\usepackage{soul}

\begin{document}

\title{Morphology, Polarization Patterns, Compression, and Entropy Production \\ in Phase-Separating Active Dumbbell Systems}

\author{Lucio Mauro Carenza}
\affiliation{Dipartimento  Interateneo di  Fisica,  Universit\`a  degli  Studi  di  Bari \\ and INFN, Sezione  di  Bari, via  Amendola  173,  Bari,  I-70126,  Italy}

\author{Claudio Basilio Caporusso}
\affiliation{Dipartimento  Interateneo di  Fisica,  Universit\`a  degli  Studi  di  Bari \\ and INFN, Sezione  di  Bari, via  Amendola  173,  Bari,  I-70126,  Italy}

\author{Pasquale Digregorio}
\affiliation{Dipartimento  Interateneo di  Fisica,  Universit\`a  degli  Studi  di  Bari \\ and INFN, Sezione  di  Bari, via  Amendola  173,  Bari,  I-70126,  Italy}

\author{Antonio Suma}
\affiliation{Dipartimento  Interateneo di  Fisica,  Universit\`a  degli  Studi  di  Bari \\ and INFN, Sezione  di  Bari, via  Amendola  173,  Bari,  I-70126,  Italy}

\author{Giuseppe Gonnella}
\affiliation{Dipartimento  Interateneo di  Fisica,  Universit\`a  degli  Studi  di  Bari \\ and INFN, Sezione  di  Bari, via  Amendola  173,  Bari,  I-70126,  Italy}

\author{Massimiliano Semeraro}
\affiliation{Dipartimento  Interateneo di  Fisica,  Universit\`a  degli  Studi  di  Bari \\ and INFN, Sezione  di  Bari, via  Amendola  173,  Bari,  I-70126,  Italy}

\begin{abstract}
Polar patterns and topological defects are ubiquitous in active matter. In this paper, we study a paradigmatic polar active dumbbell system through numerical simulations, 
to clarify how polar patterns and defects emerge and shape evolution. 
We focus on the interplay between these patterns and morphology, domain growth, irreversibility, and compressibility, tuned by dumbbell rigidity and interaction strength. 
Our results show that, when separated through MIPS, dumbbells with softer interactions can slide one relative to each other and compress more easily, producing blurred hexatic patterns, polarization patterns extended across entire hexatically varied domains, and stronger compression effects. 
Analysis of isolated domains reveals the consistent presence of inward-pointing topological defects that drive cluster compression and generate non-trivial density profiles, whose magnitude and extension are ruled by the rigidity of the pairwise potential. 
Investigation of entropy production reveals instead that clusters hosting an aster/spiral defect are characterized by a flat/increasing entropy profile mirroring the underlying polarization structure, thus suggesting an alternative avenue to distinguish topological defects on thermodynamical grounds. 
Overall, our study highlights how interaction strength and defect-compression interplay affect cluster evolution in particle-based active models, and also provides connections with recent studies of continuum polar active field models.
\end{abstract}

\maketitle

\section{Introduction}
\label{sec:intro}

{\it Active matter} defines a class of biologically inspired systems whose single constituents are able to transform stored or ambient energy into self-propulsion \cite{marchetti2013, ramaswamy2010, bechinger2016, needleman2017, fodor2018}. Physics, biology, and material science provide countless examples, ranging from macroscopic schools of fishes, flocks of birds, and herds of mammals \cite{parrish1997, reynolds1998, vicsek2012} to microscopic colonies of cells and bacteria \cite{berg2004, bi2016, maggi2023} and synthetic self-propelled colloids \cite{paxton2004, golestanian2007, walther2008, palacci2013}. The mere introduction of a self-propulsion mechanism produces distinctive phenomena with no counterpart in passive systems, as collective motion \cite{zhang2010, vicsek2012, sanchez2012, elgeti2015, hakim2017}, motility-induced phase separation (MIPS) \cite{buttinoni2013, cates2015, bialke2015}  and dynamical phase transitions \cite{gradenigo2019, semeraro2023}, which attracted great research interest in recent years and still pose major challenges \cite{gompper2025}.

\begin{figure}[h!]
\centering
\includegraphics[width=0.8\columnwidth]{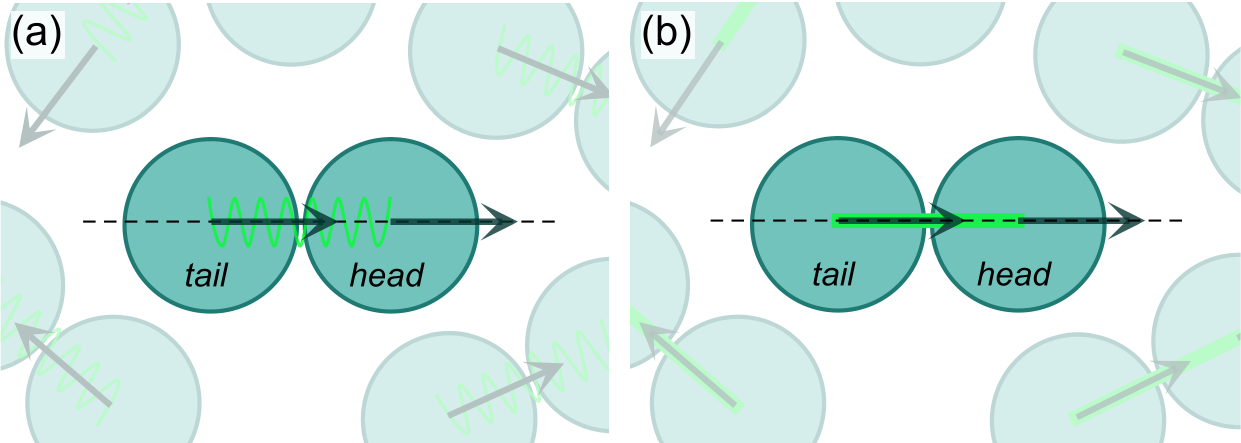}
\caption{{\textbf{Sketch of typical active dumbbell systems.} Each dumbbell is composed of two beads, tail and head, connected together. Self-propulsion (arrows) acts on each bead along the axis of its dumbbell (dashed line) with constant magnitude $F_a$ in the tail--head direction (Equation~\eqref{eq:act_force}). All beads interact through a shifted and truncated pairwise potential (Equation~\eqref{eq:pot_n}) of Lennard--Jones (\textbf{a})~or Mie (\textbf{b}) kind. In the first case, head and tail are kept together through a FENE force (Equation~\eqref{eq:fene}, green zig-zag line), thus their distance can slightly oscillate. In the second, a RATTLE scheme, which keeps head--tail distance fixed, is instead implemented (thick green line).}}
\label{fig:fig1}
\end{figure}

In numerous active systems, minimal units show peculiar tail--head asymmetries that make them intrinsically {\it {polar}} \cite{toner2005, marchetti2013, needleman2017, chate2020, paoluzzi2024}. Relevant instances are motile cells and \mbox{bacteria \cite{lopez2015, laang2024}}, active filaments \cite{schaller2010}, and synthetic self-propelled rods \cite{bar2020}. Here, local orientation is either due to explicit alignment interactions among nearby units, as in flocking models \cite{toner2005, cavagna2014, solon2015}, or to intrinsic shape properties, as in active rods \cite{ginelli2010}. It is therefore natural to consider models able to capture the persistent motion along a locally preferred direction of these units. A paradigmatic particle-based polar model is that of {\it {active dumbbells}} \cite{sum2014}, in which each unit features two connected beads, {\it {tail}} and {\it {head}}, and self-propels in the tail--head direction (see Figure~\ref{fig:fig1}). Several morphological and dynamical aspects of active dumbbell systems have already been studied \cite{joyeux2016, mandal2017, clopes2022}. In particular, MIPS is well established \cite{gonnella2014, tung2016, cugliandolo2017, venkatareddy2023, caporusso2024b, digregorio2025}, and phase diagrams were determined in two \cite{petrelli2018} and \mbox{three \cite{caporusso2024} dimensions}.

Ordered patterns in polar systems can be disrupted by the emergence of {\it {topological defects}} \cite{angheluta2025} as {\it {vortices}}, {\it {asters}}, or {\it {spirals}} \cite{kruse2004, elgeti2011, mondal2023, wang2023}, which significantly influence the dynamics of polar systems. These indeed promote and sustain large-scale coherent states around their point cores, thus bridging small- and large-scale behavior and driving macroscopic self-organization \cite{schaller2013, bricard2015, sergerer2015, guillamat2022, skogvoll2023, laang2024, ho2024}. Furthermore, topological defects can play a pivotal role in phase ordering kinetics. In this regard, it was recently shown that spirals and aster defects indeed drive an enhanced domain growth both at the particle \cite{caporusso2024b, carenza2025} and field \cite{semeraro2025} level. Interestingly, in \cite{semeraro2025}, this was interpreted in terms of the interplay between inward-pointing defects and {\it {compression}} of the density field.

From a thermodynamical point of view, continuous energy transformation into directed motion makes active systems inherently out-of-equilibrium \cite{dabelow2019, fodor2022}, thus letting them naturally break detailed balance. Quantification of irreversibility due to activity is provided by {\it {entropy production}} \cite{fodor2016, mandal2017b, pietzonka2017, grandpre2019, bebon2025}, and its investigation is thus central to fully characterize active systems. Recent studies \cite{crosato2019, borthne2020, caprini2023, beer2025} have shown that topological defects, phase separation, and large-scale flows in polar active systems can strongly modulate local and global entropy production, offering a new thermodynamical perspective on pattern formation in active matter. In particular, in \cite{semeraro2024}, it was recently shown that entropy distribution displays peculiar tail structures related to the motion of active particles in regions of low local order, i.e., close to vacancy defects, where
they can either move towards empty regions or bounce
into other particles. Additional insights are provided by the investigation of phase-separating hot and cold \cite{venkatareddy2023} and confined \cite{hutter2022} dumbbells.

In this paper, we numerically investigate clustering phenomena due to MIPS in polar systems of two-dimensional active dumbbells. Our aims are two-fold. First, we want to investigate the interplay between polarization patterns on the one side and topological defects and compression in clusters on the other. Second, we want to clarify how the overall emerging phenomenology is affected by varying dumbbell rigidity and interactions. These are, respectively, tuned by letting the distance between beads in each dumbbell slightly oscillate (or not) and by letting beads interact either through a soft- or a hard-core repulsive potential. To these aims, we first provide a comparison between clustering dynamics, morphology, and domain growth in the soft- and hard-core cases, and then concentrate on the internal structure of individual clusters in which a topological defect is clearly appreciable. Contact with continuous frameworks is made by employing numerical coarse-graining procedures to obtain density and polarization fields. 

Our investigation reveals key differences arising from contrasting mechanisms: hard-core interactions favor rigid interlocking of dumbbells, whereas soft-core ones facilitate bead sliding and compression. As a consequence, softer interactions give rise to blurred hexatic patterns, polarization patterns extended across entire hexatically varied domains, and stronger compression effects, the latter also being present, although less evident, with harder interactions. In particular, isolated clusters are found to host inward-pointing topological defects driving domain compression, thus resulting in non-trivial density profiles, whose magnitude increases with domain dimension and interaction softness. Overall, these results give us the additional opportunity to draw connections with the framework of \cite{semeraro2025}, so as to preliminarily assess whether the top-down description proposed there can serve as a representative picture of phase separation in active particle systems. A further bead-wise investigation on entropy production revealed that dumbbells close to grain boundaries between hexatic patches display larger entropy values due to their reduced, yet not zeroed, mobility in these regions. Moreover, clusters hosting an aster (spiral) defect display a flat (increasing) entropy profile stemming from the frozen (rotational) dynamics imposed by their underlying polarization structure, thus suggesting an alternative way to identify topological defects through the irreversibility they generate.

The paper is structured as follows. In Section~\ref{sec:model_methods}, we present the active dumbbell models we employ and detail the numerical approach we implement. In Section~\ref{sec:overview}, we provide an overview of the system behavior that emphasizes differences due to the adoption of different interaction rules. In Section~\ref{sec:dens_pol_prof}, we investigate density and polarization fields in isolated clusters, and the role played by topological defects. In Section~\ref{sec:ep}, we investigate entropy production of clusters and aggregates. Finally, in Section~\ref{sec:conclusions}, we draw the conclusions of our investigation.

\section{Model and Methods}
\label{sec:model_methods}

The present section is devoted to Model and Methods of our investigation. In \mbox{Section~\ref{sec:model}} we detail the active dumbbells system we consider, specifying the dumbbells setting and interaction potentials adopted. In Section~\ref{sec:num_met} we instead detail the numerical methods we adopted.

\subsection{Model}
\label{sec:model}

We study two-dimensional configurations with $N$ identical active dumbbells, a sketch of which is provided by Figure~\ref{fig:fig1}. Each dumbbell consists of a diatomic molecule made of two identical circular beads, tail and head, of diameter $\sigma_d$ and mass $m_d$, thus resulting in $2N$ beads. Beads are orderly labeled as $i=1,\ldots,2N$, with odd (even) indices assigned to tail (head) beads, so that consecutive odd--even indices apply to the same dumbbell. The dynamics of the $i$-th bead is ruled by the following Langevin equation: 

\begin{equation}
m_d\ddot{\bm{r}}_i(t)=-\gamma_d\dot{\bm{r}}_i(t)-\nabla_iU(\{\bm{r}_i(t)\})+\bm{F}_{link}(\bm{r}_{i,i\pm1})+\bm{F}_{a,i}(t)+\sqrt{2k_BT\gamma_d}~\bm{\xi}_i(t)~,
\label{eq:lang_eq_bead}
\end{equation}
where $\bm{r}_i(t)$ denotes the bead position, $\nabla_i\equiv \partial_{\bm{r}_i}$, $\gamma_d,~T$ respectively, are the friction coefficient and temperature of the solvent the dumbbells are immersed in, and $k_B$ is the Boltzmann constant. The symbol $\bm{\xi}_i(t)$ denotes one of $2N$ independent Gaussian white noises that model the action of the thermal bath on each bead. These feature vanishing mean and delta correlation, i.e.,
\begin{equation}
\braket{\xi_{i,a}(t)}=0\qquad\text{and}\qquad\braket{\xi_{i,a}(t)\xi_{j,b}(s)}=\delta_{ij}\delta_{ab}\delta(t-s)~,
\end{equation}
with $a,b=x,y$ spatial coordinates, $\delta_{ij},~\delta_{ab}$ Kronecker deltas, and $\delta(t-s)$ a Dirac delta.

The potential $U(\{\bm{r}_i(t)\})$ accounts for overall repulsive interactions. For any two beads, say the $i$-th and the $j$-th, they are ruled by
\begin{equation}
U(r_{ij}(t))=\left\{4\epsilon\left[\left(\frac{\sigma}{r_{ij}(t)}\right)^{2n}-\left(\frac{\sigma}{r_{ij}(t)}\right)^n\right]+\epsilon\right\}\Theta(r_c-r_{ij}(t))~,
\label{eq:pot_n}
\end{equation}
where $\sigma$ and $\epsilon$, respectively, denote the length and energy scale of the potential, $r_{ij}=|\bm{r}_i(t)-\bm{r}_j(t)|$ and $\Theta(r_c-r_{ij}(t))$ is a Heaviside function (we assume the convention $\Theta(0)=0$). The $\epsilon$ summand shifts the potential in such a way that its minimum is at $0$. The specific choice of $\sigma$ depends on $n$ and will be detailed in a moment. The overall interaction potential $U(\{\bm{r}_i(t)\})$ is obtained as the sum of all instances of Equation~\eqref{eq:pot_n} for all couples of beads, both within the same dumbbell or in different ones. Concerning specific instances of Equation~\eqref{eq:pot_n}, we consider two different possibilities:
\begin{itemize}
\item	$n=6$: the non-vanishing part of Equation~\eqref{eq:pot_n} reduces to a purely repulsive Lennard--Jones (LJ) potential \cite{jones1931}, which yields a soft repulsive core and allows for appreciable bead overlap and sliding. In this case we set $r_c=\sigma_d$ and $2^{1/6} \sigma=\sigma_d$.
\item	$n=32$: the non-vanishing part reduces instead to the Mie potential \cite{mie1903}, which provides a harder repulsive core, and therefore limits bead overlap.  As typical for harder interactions \cite{cugliandolo2017}, in this case, we instead set $r_c=\sigma_d$ and $\sigma=\sigma_d$.
\end{itemize}

Tail and head within each dumbbell are connected through the action of the force $\bm{F}_{link}(\bm{r}_{i,i\pm1})$. Dumbbell rigidity is tuned by letting $\bm{F}_{link}(\bm{r}_{i,i\pm1})$ take different 
expressions depending on the interaction potential we consider:
\begin{itemize}
\item In the LJ case, we promote bead overlapping; thus, $\bm{F}_{link}(\bm{r}_{i,i\pm1})$ is set as a finite extensible non-linear elastic (FENE) force \cite{warner1972}, which at the same time avoids unlimited bead separation (see Figure~\ref{fig:fig1}(a)). In symbols, FENE force is defined as
\begin{equation}
\bm{F}_{FENE}(\bm{r}_{i,i\pm1})=\frac{k \bm{r}_{i,i\pm1}(t)}{1-(r_{i,i\pm1}(t)/r_0)^2}~,
\label{eq:fene}
\end{equation}
where $k$ is an elastic constant, $r_0$ is the maximum distance between beads, and $\bm{r}_{i,i\pm1}(t)=\bm{r}_i(t)-\bm{r}_{i\pm1}(t)$, $r_{i,i\pm1}(t)=|\bm{r}_{i,i\pm1}(t)|$, with $i+1$ ($i-1$) when the $i$-the bead is tail (head).

\item In the Mie case, we instead promote dumbbell rigidity; thus, we keep the bead distance fixed at $\sigma_d$ by adopting the RATTLE scheme \cite{andersen1983} (see Figure~\ref{fig:fig1}b). This is equivalent to identifying $\bm{F}_{link}(\bm{r}_{i,i\pm1})$ with a force that takes into account
holonomic constraints. 
\end{itemize}

Finally, self-propulsion acts on each bead along the axis of its dumbbell with constant magnitude $F_a$ in the tail--head direction (see Figure~\ref{fig:fig1}). More specifically, for each dumbbell we define a unit vector $\hat{n}_{i,i+1}(t)$ connecting the centers of the tail and head bead and apply the active force
\begin{equation}
\bm{F}_{a,i}(t)=\bm{F}_{a,i+i}(t)=F_a\hat{n}_{i,i+1}(t)~
\label{eq:act_force}
\end{equation}
at the center of the head and tail beads. Although having a constant modulus, $\bm{F}_{a,i}(t)$, it still shows a time-dependency as its direction varies according to the instantaneous \mbox{dumbbell orientation}.

The overall system of $N$ dumbbells evolves in a square box of side $L$ with periodic boundary conditions. We explore different system configurations by varying two adimensional control parameters, the surface fraction
\begin{equation}
\rho=\frac{N\pi\sigma_d^2}{2L^2}~,
\end{equation}
which measures the area fraction covered by all beads in the box, and the P\'eclet number
\begin{equation}
Pe=\frac{2F_a\sigma_d}{k_BT}~,
\label{eq:pe}
\end{equation}
which instead provides a comparison between the typical work performed by the active force to propel a dumbbell by its typical size and the thermal energy scale set by the bath.

\subsection{Numerical Methods}
\label{sec:num_met}

Our investigation relies on a numerical approach. We integrate the Langevin equations Equation~\eqref{eq:lang_eq_bead} using the open-source package Large-scale Atomic/Molecular Massively Parallel Simulator (LAMMPS, release 2 August 2023) \cite{plimpton1995, lammps}, employing a Velocity-Verlet scheme with a Langevin thermostat to have an NVT ensemble and periodic boundary conditions. The system parameters are fixed to $\sigma_d=1$, $m_d=1$, $\gamma_d=10$, $k_BT=0.05$, $k=30$, $r_0=1.5$, and $\epsilon=1$, thus assuring the overdamped limit. Accordingly, $\sigma_d$, $m_d$, and $\epsilon$ define the reduced units of length, mass, and energy \cite{allen2017}, which also make the molecular dynamics time unit $\tau_{MD}=\sqrt{m\sigma_d^2/\epsilon}$ unitary. The integration timestep is chosen as $\Delta t=10^{-2}\tau_{MD}-10^{-3}\tau_{MD}$ to guarantee numerical stability and convergence. 

We focus on $(\rho,Pe)$ regimes in which phase separation and coexistence occur. The density $\rho$ is controlled by varying $N$ between $2^{19}$ and $2^{20}$ (thus the number of dumbbells, respectively, is $2^{18}=512^2$ and $2^{19}=1024^2/2$) and adjusting the box size $L$ to match the target surface fraction $\rho$, which we vary in $[0.4,~0.6]$. The target P\'eclet numbers we explore instead are $Pe=100$ and $200$, which we obtain by setting $F_a=2.50$ and $5.00$. The additional $F_a=0.20, 0.25$, and $0.75$ value are considered to build the phase diagram in Figure~\ref{fig:figs1}.

For each $\rho$, dumbbells are initially placed in the simulation box in a homogeneous state, i.e., at random locations and with random orientation. Then, once $F_a$ is also fixed, we let the system evolve for up to $\sim$$10^6\tau_{MD}$. For each pair $(\rho, Pe)$ we run up to $10$ independent simulations on $112$ cores, for a total of up to $24$ h for each CPU.

Analyses are performed on output configurations offline. Details on each kind of analysis are provided throughout the paper.

\section{Overview of System Phenomenology: Domain Growth, Hexatic Order, Polarization Patterns, and Compression}
\label{sec:overview}

In the present section, we provide an overview of the clustering dynamics and morphology of the system. To do so, we set parameters in such a way that the homogeneous initial state tends to phase separate and form clusters (see phase diagram in Figure~\ref{fig:figs1}). While phase behavior \cite{cugliandolo2017, petrelli2018} and clustering phenomena \cite{petrelli2018, caporusso2024b} have been extensively studied for hard repulsive dumbbells, our analysis emphasizes the differences that arise when the LJ or Mie setting is employed. In the following we will use the terms cluster and domain indistinctly.
 
\subsection{Domain Growth, Hexatic Order, and Cluster Shape}
\label{sec:hex_ord}

As a first thing, we monitor the degree of hexatic order within our systems in connection with domain growth and cluster shapes. Domain growth is monitored by looking at the characteristic domain size $R(t)$, which, as typically done \cite{kendon2001, caporusso2024b}, is evaluated as the inverse of the first moment of the normalized spherically averaged structure factor, i.e., as $R(t)=\pi\int dk~S(k,t)/\int dk~kS(k,t)$. Here, the structure factor is defined as $S(\bm{k},t)=\frac{1}{2N}\sum_{i}^{2N}\sum_{j}^{2N}e^{\imath\bm{k}\cdot(\bm{r}_i(t)-\bm{r}_j(t))}$ with $\bm{k}$ wave vector in the Fourier space, while its radial symmetric average is obtained by averaging over spherical shells of thickness $k=2\pi/L$ in $\bm{k}$ space. Hexatic order is instead monitored by computing particle-wise the local hexatic order parameter $\psi_{6i}=\sum_{j=1}^{n_i}e^{\imath6\theta_{ij}}/n_i$, with $n_i$ the number of nearest neighbors of the $i$-th bead extracted from a Voronoi tessellation and $\theta_{ij}$ the angle between the segment that connects the $i$-th bead with its $j$-th neighbor and the $x$-axis. Hexatic order is visualized as proposed in \cite{bernard2014}: first, the complex values of $\psi_{6i}$ are projected onto the direction of their space average. Next, each bead is colored according to this projection. Regions with orientational order have uniform color.

\begin{figure}[t!]
\centering
\includegraphics[width=0.8\columnwidth]{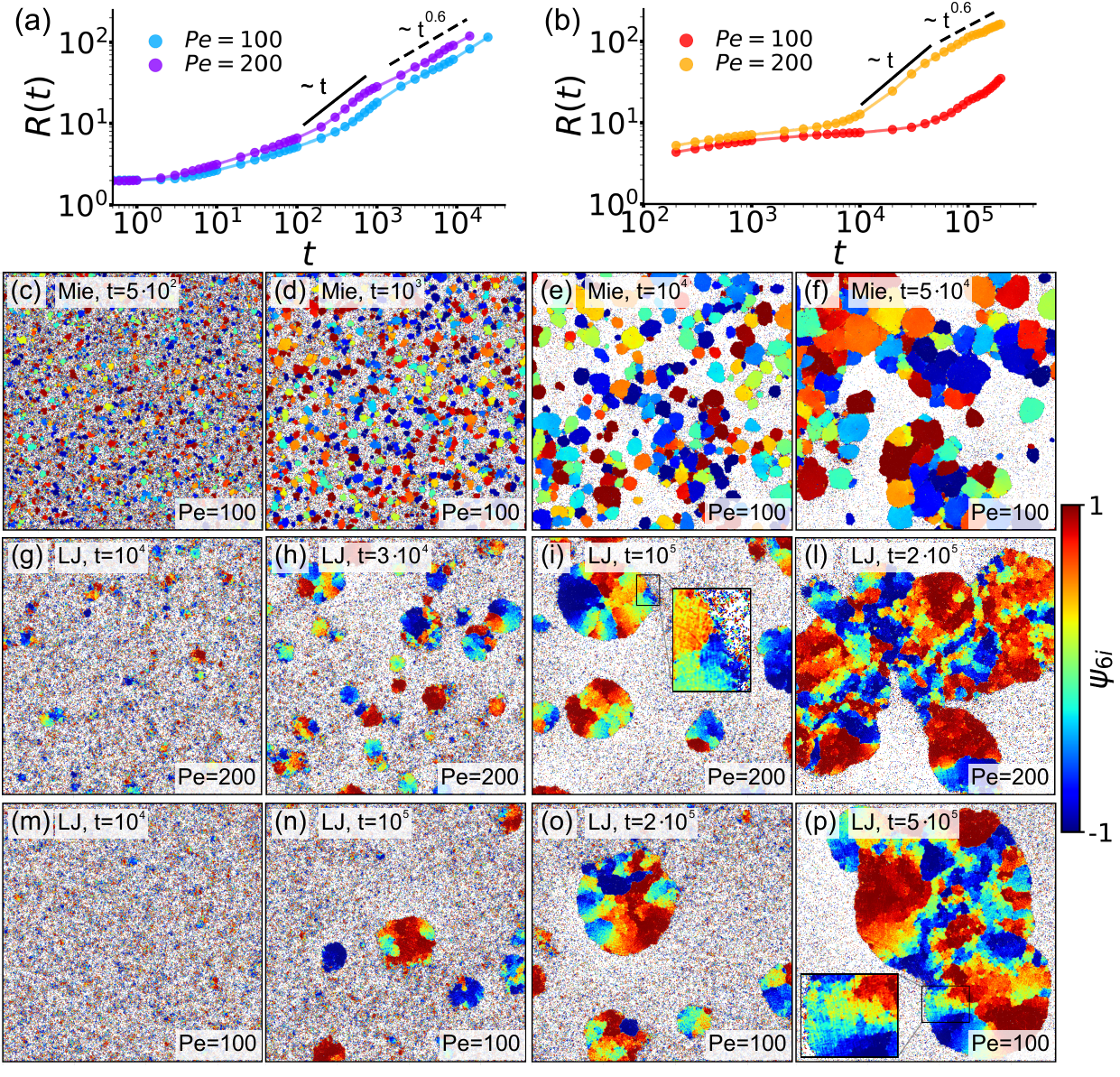}
\caption{{\textbf{Growth regimes and hexatic order.} (\textbf{a},\textbf{b}). Characteristic domain size $R(t)$ for $Pe=100,~200$ in the Mie and LJ configurations, respectively. (\textbf{c}--\textbf{p}). Snapshots of an enlarged area of the system in the Mie at $Pe=100$ (\textbf{f}--\textbf{n}), LJ at $Pe=200$ (\textbf{g}--\textbf{j}), and LJ at $Pe=100$ (\textbf{k}--\textbf{n}) configurations at $\rho\sim 0.4$. Beads are colored according to their $\psi_{6i}$ values (right bar).}}
\label{fig:fig2}
\end{figure}

In Figure~\ref{fig:fig2}a,b we report the trend of $R(t)$ at $Pe=100,200$ in the Mie and LJ configurations. As also reported in \cite{caporusso2024b}, in the Mie case, three different growth regimes can be identified: a nucleation regime during which the cluster typical length $R(t)$ increases quite slowly, a condensation--aggregation regime in which $R(t)\sim t$, and finally a translation--collision--merging regime in which $R(t)\sim t^{0.6}$. As for the LJ case, in line with~\cite{carenza2025}, Figure~\ref{fig:fig2}b reveals these regimes to be confirmed for $Pe$ large enough. For lower $Pe$ values, instead, domains still grow in size, but no power-law trend can be recognized.

Provided the activity is large enough, the evolution of the system in the Mie and LJ configurations thus seems to proceed similarly. However, at a closer look, relevant differences dictated by the different interaction rules emerge. Interestingly, all of these can be traced back to the mere fact that Mie interactions promote hard dumbbell interlock, while LJ interactions forward dumbbell sliding. Let us be more specific. 

Concerning domain growth, we observe that at comparable $Pe$ values, growth dynamics markedly differ between the Mie and the LJ cases. In the LJ configuration, the ability of dumbbells to slide on each other, together with weaker interlock, produces an overall slowdown in the coarsening processes, so that domain growth takes place on longer timescales than in the Mie case. According to Figure~\ref{fig:fig2}a,b, and considering the times at which the growth regimes change, this slowdown amounts to roughly an order of magnitude. 

This fundamental difference in kinetics underlies a series of structural distinctions between the two configurations at subsequent evolution stages. These differences are highlighted in the remainder panels of Figure~\ref{fig:fig2}, where we report snapshots of enlarged areas of the system in the Mie (panels c--f) and LJ (panels g--j and k--n) configurations, with beads colored according to their $\psi_{6i}$ values. Extraction times are chosen in such a way as to illustrate how dumbbells and clusters arrange during different growth stages. For the Mie case, here and in the following, we analyze just the case $Pe=100$, as an overall similar scenario is observed also at $Pe=200$ (see for example \cite{petrelli2018}). Prompted by Figure~\ref{fig:fig2}b, for the LJ case, we instead preliminarily consider both the cases $Pe=100$ and $200$ as a way to illustrate how this (up to our knowledge) less characterized setting behaves at different activity strengths. During the nucleation regime, Mie configurations are characterized by the rapid formation of cluster cores (see Section~\ref{sec:pol_prof} for details) on which then additional dumbbells rigidly interlock, thus generating hexatically ordered domains (see Figure~\ref{fig:fig2}c,d). By contrast, in the LJ configurations, dumbbells still interlock, but can also more easily slide. As a consequence regions with variegated hexatic order, or {\it {rainbow patterns}}, emerge (see insets in Figure~\ref{fig:fig2}i,n). Interestingly, this initial difference in local organization impacts the domain structures observed at later times. As the system enters the translation–collision–merging regime, these structures, and their interplay, determine how larger aggregates build up. In the Mie case, collisions between hexatically ordered domains lead to polycrystalline structures composed of distinct hexatic patches separated by grain boundaries (see Figure~\ref{fig:fig2}e,f). Instead, in the LJ case at larger $Pe$, the same sliding mechanism delaying nucleation also blurs boundaries between merging clusters. As a consequence, rainbow patterns are observed also at transient interfaces, which become generally less sharply defined (see Figure~\ref{fig:fig2}i,j,m,n). This behavior is clearly illustrated in Figure~\ref{fig:fig2}n, where the central large cluster has merged with two smaller ones (top left and bottom right), and is further evidenced by its inset.

Because of these different microscopic processes, the cluster morphology evolves differently. In Mie configurations, until the largest simulation times we explored, clusters tend to remain irregular and elongated as the rigid interlock between dumbbells and smaller clusters prevents shape relaxation. However, we report that in \cite{caporusso2024b}, it is speculated that configurations with a single orientational order should be the truly stable ones. These patchwork patterns should likely be metastable, arising from collisions between clusters of different hexatic orders whose interfaces persist due to insufficient simulation time. Yet, no direct observation of these rearrangements is yet available due to the heavy computational cost required. Conversely, in LJ configurations, clusters are progressively reshaped by dumbbell sliding, which promotes the formation of rounder aggregates (see Figure~\ref{fig:fig2}i,m). This effect is promoted at lower $Pe$, as the lower activity push allows beads to rearrange more easily. In both settings, however, no bubbles are observed, as instead for circular active particle systems \cite{caporusso2020} and scalar continuum models \cite{tjhung2018}.

While in both LJ cases we treated the qualitative phenomenology is very similar, without loss of generality, in the following, we focus on the case at $Pe=200$ as it generates an $R(t)$ curve with well separated regimes more similar to the ones from the Mie case.

\subsection{Local Polarization Density Field and Polarization Patterns}
\label{sec:pol_field}

As mentioned in the introduction, one of the reasons we consider dumbbell systems is the inherent polar character of their minimal constituents. In order to monitor how dumbbells' orientation contributes to the overall polarization of the system on scales larger than their dimensions, we perform a coarse-graining procedure. More specifically, we determine a coarse-grained local polarization density field $\vec p_i$ in two steps: first, we divide the simulation box into small squares of side $\sigma_{cg}$; next, for each square, we compute $\vec p_i$, where the subscript $i$ remarks its local character, as the vector sum of all unit vectors belonging to dumbbells within the square divided by the square surface $\sigma_{cg}^2$. We tested different $\sigma_{cg}=10,~15$, and $20\sigma_d$ at the same time larger than dumbbell dimensions and lower than the typical size of aggregates, finding in all cases consistent results. In the following, we concentrate on the case $\sigma_{cg}=10\sigma_{d}$, which provides the best representation resolution.

\begin{figure}[t!]
\centering
\includegraphics[width=0.8\columnwidth]{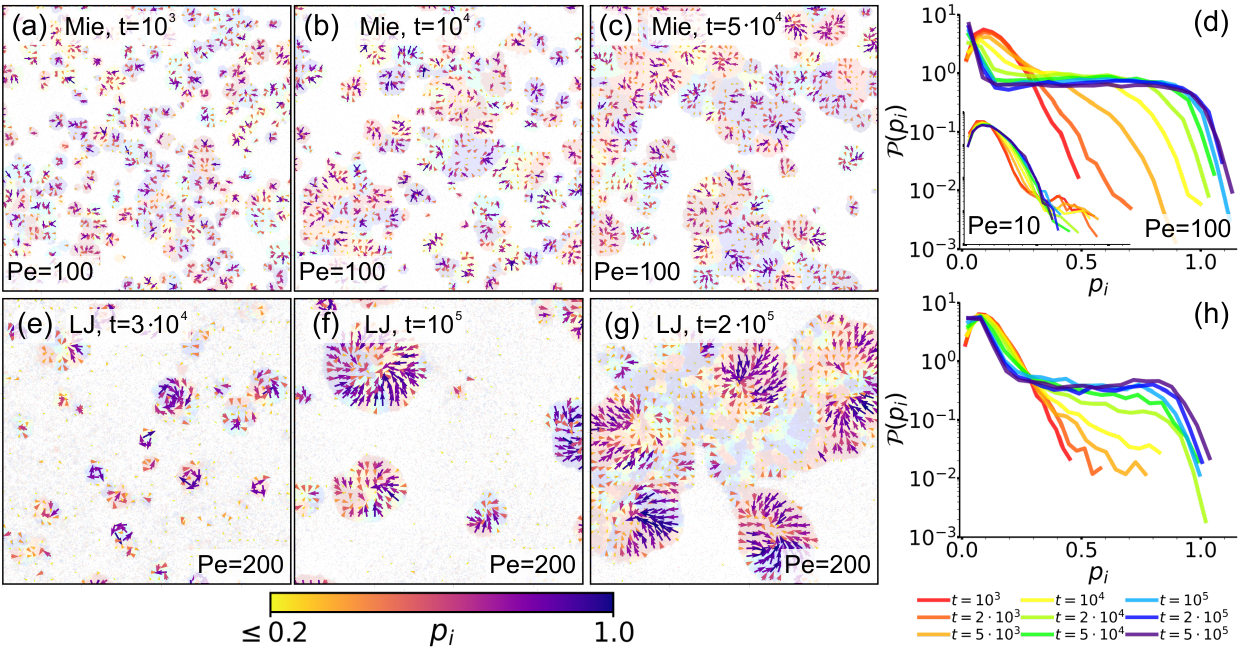}
\caption{{\textbf{The local polarization density field.} (\textbf{a}--\textbf{c},\textbf{e}--\textbf{g}). Local polarization density field $\vec p_i$ with $\sigma_{cg}=10\sigma_d$ in an enlarged area of the system at $Pe=100$ in the Mie and at $Pe=200$ in the LJ configuration at $\rho\sim 0.4$, respectively. Extraction times and enlarged areas are as in Figure~\ref{fig:fig2} for the three respective larger times. The field is represented as arrows colored according to their magnitude $p_i=|\vec p_i|$ (bottom bar). Arrows such that $ p_i<0.2$ are removed. Backgrounds report corresponding hexatic snapshots from Figure~\ref{fig:fig2}. (\textbf{d},\textbf{h}). Distribution of the field magnitude $p_i$ in the Mie and LJ configurations, respectively. The inset in (\textbf{d}) reports the distributions obtained at $Pe=10$ with axes as in the main panel. All curves are generated at increasing times (bottom legend), collecting data from $10$ independent simulation runs.}}
\label{fig:fig3}
\end{figure}

In Figure~\ref{fig:fig3}a--c,e--g, we respectively report the output field $\vec p_i$ for the Mie and LJ cases. Extraction times and enlarged areas, as in Figure~\ref{fig:fig2}, are shown for the three respective largest times. For clarity, in the background we report the corresponding hexatic snapshots from Figure~\ref{fig:fig2}. These panels clearly illustrate the formation of peculiar polarization patterns. As discussed in Section~\ref{sec:pol_prof}, the initial nucleation of small domains naturally fosters the formation of spiraling patterns in both Mie and LJ configurations. However, also in this case, there are interesting differences due to the interaction rule. In Section~\ref{sec:hex_ord}, we mentioned that for Mie configurations, small nucleated domains initially grow by dumbbell aggregation into hexatically ordered ones, and then, by collision and merging, these give rise to hexatically patched larger clusters. In terms of polarization patterns, Figure~\ref{fig:fig3}b,c show that each of these patches tends to retain its polarization structure  (see also Figure~\ref{fig:figs2} for additional snapshots over the entire system and {Figure~\ref{fig:fig5}}g--i for closer enlargements). This therefore suggests that the structural integrity of these larger clusters should be mainly ensured by strong interlock between patches. On the contrary, in LJ configurations, we observe the appearance of rainbow patterns due to dumbbell sliding (see also Figure~\ref{fig:figs3} for additional snapshots over the entire system). While this sliding mechanism hinders hexatic order, Figure~\ref{fig:fig3}e,f show it to instead favor the formation of unified polarization patterns (see also Figure~\ref{fig:fig5}j--m). In other words, instead of observing many separated patterns, we observe just a common one that interests the entire cluster. In addition, already at this stage but also in the translation--collision--merging regime, we observe that at cluster periphery the system is strongly polarized (see also Figure~\ref{fig:fig5}d--f), with the $\vec p_i$ field generally pointing inward. This therefore suggests that now structural integrity is additionally fostered by these inward-pointing dumbbells.

Complementary quantitative information is provided by Figure~\ref{fig:fig3}d,h. Here we, respectively, report the distribution $\mathcal{P}(p_i)$ of the polarization magnitude $p_i=|\vec p _i|$ in the Mie and LJ cases at different times. These curves reflect the phenomenology and differences discussed until this point. In particular, at small times the distributions in both cases show a similar peak at similar small values, in turn similar to the peak observed at all times at low $Pe$ (see inset in Figure~\ref{fig:fig3}d), for which clustering is hindered. This clearly signals that dumbbells are still moving essentially freely when clusters, and thus polarization patterns, have not yet formed. As soon as they do, we instead observe differences. In the Mie case, larger values of $p_i$ become rapidly more represented, until the distribution essentially reaches a stationary form. This, at the same time, reflects the rapid domain growth observed in Figure~\ref{fig:fig2}a and also the late-time merging mechanism, in which different hexatic patterns interlock without varying much their internal polarization structure. In the LJ case we instead observe that larger $p_i$ values become more represented at larger times. These findings reflect the slower domain growth observed in Figure~\ref{fig:fig2}b as well as a slower rearranging mechanism of clusters in rounder shapes with unified patterns, only at the end of which $p_i$ increases appreciably.

\subsection{Local Density Field and Compression}
\label{sec:dens_field}

To conclude our overview, we monitor how different interaction rules affect bead compression, as well as its interplay with underlying cluster polarization structures. To do so, similarly to Section~\ref{sec:pol_field}, we perform a coarse-graining procedure to obtain a local density field $\phi_i$. In each square of side $\sigma_{cg}$, the scalar field $\phi_i$, with the subscript $i$ still denoting local character, is now estimated as the surface fraction occupied by all beads within the square.

\begin{figure}[t!]
\centering
\includegraphics[width=0.8\columnwidth]{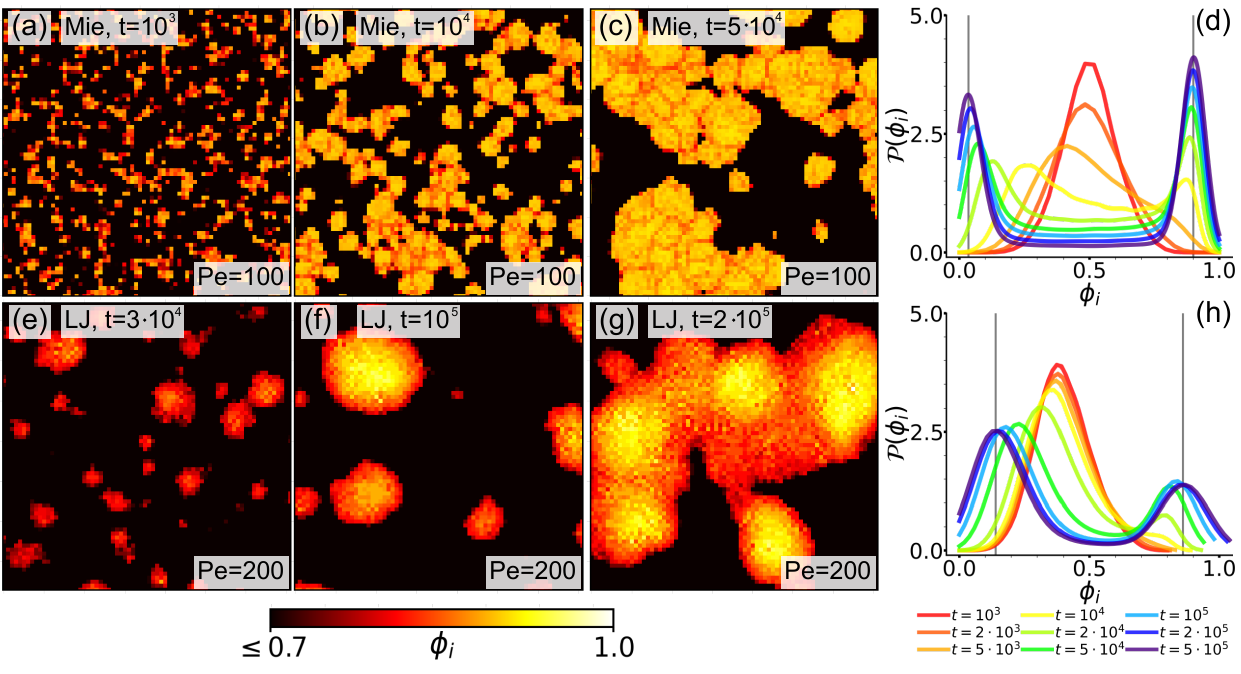}
\caption{{\textbf{The local density field.} (\textbf{a}--\textbf{c},\textbf{e}--\textbf{g}). Local density field $\phi_i$ with $\sigma_{cg}=10\sigma_d$ in an enlarged area of the system at $Pe=100$ in the Mie and at $Pe=200$ in the LJ configuration at $\rho\sim 0.4$, respectively. Extraction times and enlarged areas are as in Figure~\ref{fig:fig3}. The field is colored according to the bottom bar. (\textbf{d},\textbf{h}). Distribution of the local density field $\phi_i$ in the Mie and LJ configurations, respectively. Curves are generated at increasing times (bottom legend), collecting data from $10$ independent simulation runs. Vertical lines highlight the location of $\phi_{low}$ and $\phi_{high}$ at the largest time sampled.}}
\label{fig:fig4}
\end{figure}

Figure~\ref{fig:fig4} reports snapshots of the density field $\phi_i$. Let us comment on the Mie case first. From the snapshots in the top row, we observe that patched structures are still recognizable. The grain boundaries between hexatic patches are in fact characterized by slightly lower $\phi_i$ values (red pixels within large clusters in Figure~\ref{fig:fig4}b,c) due to a strong yet imprecise interlock between patches, resulting in a lower local density. This is reminiscent of the grain boundaries observed in systems of rigid disks, characterized by chains of disclination and dislocation defects \cite{digreriorio2022}. As for density values well within patches, at a first look, Figure~\ref{fig:fig4}'s top row suggests $\phi_i$ taking a constant value $\sim $0.9 with slight oscillations around it. However, a closer inspection (see Figure~\ref{fig:fig6}a) reveals that within patches $\phi_i$ values slightly increase from boundary to core, thus mirroring the underlying inward-pointing polarization structures described in Section~\ref{sec:pol_field}. This observation therefore points towards a slight dumbbell compression, which, however, is still hindered by the action of the hard repulsive \mbox{Mie potential}.

The snapshots in the bottom row of Figure~\ref{fig:fig4} reveal this phenomenology to be enhanced in the LJ case, where bead repulsion is less intense. We in fact observe that, as evolution progresses, larger clusters form with distinct density patterns. These are now clearly characterized by $\phi_i$ values decreasing from core to periphery (density profiles in isolated clusters are discussed in Section~\ref{sec:dens_prof}). Moreover, differently from the Mie case, due to the appearance of rainbow patterns, no distinct separation between different regions of the clusters is observed. Rather, here we can clearly appreciate the interesting interplay between polarization and density hinted by the Mie case. Comparing Figure~\ref{fig:fig3}e--g and density patterns reported here, we observe that, in the first place, they both interest the entire cluster in a unified manner. In addition, the peculiar structure of density patterns manifestly mirrors these underlying inward-pointing polarization structures, which, therefore, in addition to fostering structural integrity, also produce an effective compression. 

A more quantitative assessment is provided by Figure~\ref{fig:fig4}d,g. Here we, respectively, report the distribution $\mathcal{P}(\phi_i)$ in the Mie and LJ cases at different times. In order to increase smoothness, we performed a kernel density estimation procedure \cite{parzen1962}. In both cases we observe the appearance of a stable two-peak structure typical of a system undergoing phase separation. In the Mie case, after a transient regime, these peaks are located at $\phi_{low}\sim 0.04$ and $\phi_{high}\sim 0.9$, with the $\phi_{high}$ peak slightly higher, and are characterized by a reduced width. We remark that, according to \cite{digregorio2019}, phase separation at large $Pe$ in dumbbell systems generally occurs at lower activity levels than in active disk ones. In line with our above observations, the reduced width of the high-density peak, which is located at a value compatible with the close packing density $\phi_{cp}\sim 0.907$ \cite{chaikin1995}, at the same time signals that density values in the dense phase indeed concentrate around the most probable value $\phi_{high}$ and that, as $\phi_i>\phi_{cp}$ instances, although less represented, are yet present, a slight compression effect is indeed in action. In the LJ case, this overall scenario emerges more clearly. At the largest time we sampled, peaks are located at $\phi_{low}\sim 0.14$ and $\phi_{high}\sim 0.86$, displaying a much larger width, and the $\phi_{low}$ peak is now higher. The new value of $\phi_{low}$, together with larger peak widths, mirrors the significant bead compression allowed in this configuration, which additionally generates a larger spread in $\phi_i$ values (see Figure~\ref{fig:fig4}f,g). The lower $\phi_{high}$ value instead accounts for lower density values observed at cluster boundaries. Dumbbells are in fact more numerous in these regions than in areas around cores. Interestingly, its larger width, which is responsible for a reduced peak height, marks a more nourished presence of $\phi_i>\phi_{cp}$ instances, with a few $\phi_i\sim1$ occurrences.

\section{Local Density and Polarization Density Fields in Isolated Clusters}
\label{sec:dens_pol_prof}

The overview presented in Section~\ref{sec:overview} on the one side revealed an interesting connection between polarization patterns and density structures, especially when soft-core interactions are considered, while on the other, it gives the opportunity to highlight numerous parallelisms with the overall framework and phenomenology of the continuous model introduced in \cite{semeraro2025}. There, we recently showed that in a phase-separating continuum polar active model, advective effects drive the formation of domains hosting topological defects, and that compressive effects promoted by inward-pointing defects lead to an enhanced domain growth with rounder domains. The most striking similarities emerged up to this point are two: overall configurations from Figure~\ref{fig:fig2} are very similar to the ones recently observed in \cite{semeraro2025} (cfr. Figure 1 in \cite{semeraro2025}); the curves for the characteristic domain size are qualitatively similar in the two cases, both showing a regime $R(t)\sim t^{0.6}$ (compare Figure~\ref{fig:fig2}a,b with Figure 2 in \cite{semeraro2025}). These similarities become more marked
when the softer potential \mbox{is considered}.

Motivated by these observations, in the present section we analyze our systems more in depth to better emphasize the interplay between topological defects and cluster compression, and possibly provide qualitative support to the claim that our top--down description from \cite{semeraro2025} may serve as a representative picture of phase separation in active particle systems. To achieve this goal, we focus on isolated clusters with round shapes. Cluster identification is performed using the Python-implemented DBSCAN \mbox{algorithm \cite{dbscan}} (scikit-learn 1.7.2, September 2025): we consider two beads as being part of the same cluster if their distance is less than $1.1\sigma_d$ and fix to $3$ the minimum number of beads that form a cluster. Round clusters are instead identified based on their inertial tensors as the ones whose eccentricity is less than $\sim$0.6. For each selected cluster, density profiles are obtained by averaging the density field over concentric annuli of thickness $6\sigma_d$ sharing the center of mass as a common center. 

\subsection{Topological Defects}
\label{sec:pol_prof}

\begin{figure}[t!]
\centering
\includegraphics[width=0.8\columnwidth]{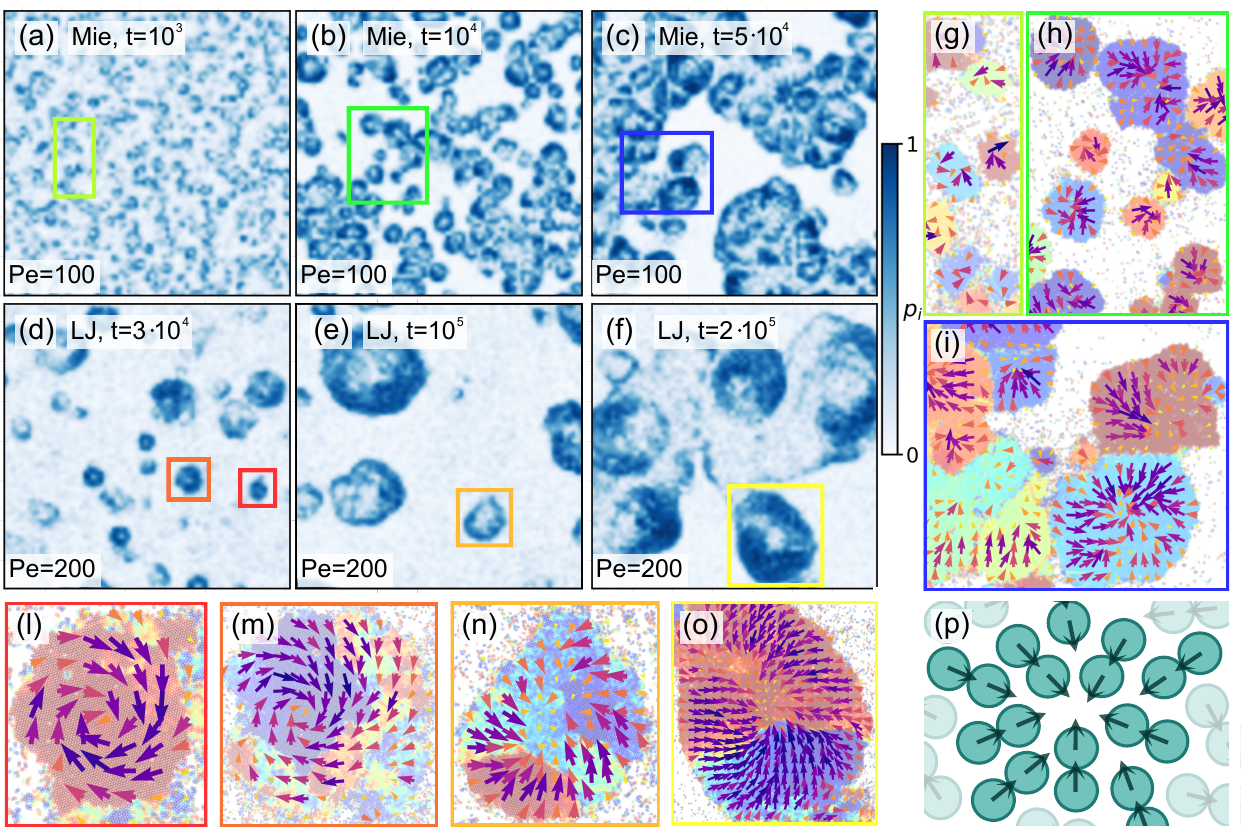}
\caption{{\textbf{Topological defects in isolated clusters.} (\textbf{a}--\textbf{f}). Local polarization density magnitude field $p_i$ with $\sigma_{cg}=10\sigma_d$ in an enlarged area of the system at $Pe=100$ in the Mie and at $Pe=200$ in the LJ configurations at $\rho\sim 0.4$, respectively. Extraction times and enlarged areas are as in Figure~\ref{fig:fig3}. The field is colored according to the right bar. (\textbf{g}--\textbf{i},\textbf{j}--\textbf{m}). Representative instances of topological defects in the LJ and Mie configurations extracted in (\textbf{a}--\textbf{f}) from regions delimited by rectangles with matching colors. Overall representation is as in Figure~\ref{fig:fig3}. (\textbf{n}). Schematic depiction of the nucleation mechanism driving the formation of topological defects.}}
\label{fig:fig5}
\end{figure}

According to definition \cite{degennes1993}, topological defects are points or regions in an ordered medium where the relevant order parameter becomes ill-defined or discontinuous, making it impossible to recover an ordered pattern through any continuous transformation. In the framework of our polar systems, defect cores can be located as points (or small regions) within clusters where the field $\vec p_i$ vanishes. In order to investigate their actual presence, in Figure~\ref{fig:fig5}a--f we report the same snapshots from Section~\ref{sec:overview}, now colored from white to blue according to the local polarization density magnitude. From these, it is possible to observe that indeed both hexatic patches in the Mie case and entire rounder clusters in the LJ one are generally characterized by a strong orientational order at the periphery (strong blue coloring), which is lost in the core (white coloring). This is clearly visible in isolated clusters (see boxed areas in Figure~\ref{fig:fig5}b,d,e) but also in larger clusters in the process of merging with smaller ones having their own polarization pattern (see boxed areas in Figure~\ref{fig:fig5}c,f). 

Direct proof of the presence of defects is provided by Figure~\ref{fig:fig5}g--m, which report close enlargements of areas from (a) to (f) delimited by rectangles with matching colors. All these enlarged snapshots report in the background beads colored according to their $\psi_{6i}$ value, as in Figure~\ref{fig:fig2}, overlapped by their corresponding $\vec p_i$ field, which is represented as in Figure~\ref{fig:fig3}. In all cases, the appearance of topological defects with unitary topological charge, typical of polar systems \cite{angheluta2025}, can be appreciated. Figure~\ref{fig:fig5}g--i provide a clearer representation of the fact that in the Mie case, each hexatic patch retains its polarization structure, which in most cases is indeed that of a defect. As for the LJ case, Figure~\ref{fig:fig5}j--m instead prove that in isolated clusters, or even in smaller clusters which just collided with a larger one, a unified defect structure insists on rainbow patterns.

Interestingly, in both Mie and LJ configurations, we observe an abundance of inward spiral-like and aster-like defects, with vortex-like defects essentially absent. That this is a general trait of our systems is supported by visual inspection of snapshots of the entire system (see Figures~\ref{fig:figs2} and \ref{fig:figs3}). This is yet another similarity with \cite{semeraro2025}. There, we in fact showed that the number of inward spiral-like and aster-like defects has a non-monotonic trend, first increasing and then decreasing, with spirals always more numerous than asters, in turn more numerous than vortices (cfr. Figure 3 in \cite{semeraro2025}). The reason behind this similarity resides in a similar mechanism for the initial formation of small clusters containing defect cores. In \cite{semeraro2025}, defects generate after the collision of multiple small advected domains. When these collide in such a way that their respective polarization points toward a common point, which is statistically more probable, an inward spiral-like or aster-like defect emerges. When instead these collide forming closed polarization lines, which is statistically less probable, a vortex-like defect is formed. The nucleation mechanism for our dumbbells model is instead illustrated in Figure~\ref{fig:fig5}n. As already remarked in \cite{gonnella2014, petrelli2018}, nucleation starts with multiple dumbbells colliding head to head while moving towards a common point, thus generating the cluster core. Afterward, additional dumbbells segregate according to the same pattern, whose stability is ensured by large activities, as in our cases. The overall natural result of this clustering mechanism is the formation of inward-pointing defects.

\subsection{Density Profiles}
\label{sec:dens_prof}

\begin{figure}[t!]
\centering
\includegraphics[width=0.8\columnwidth]{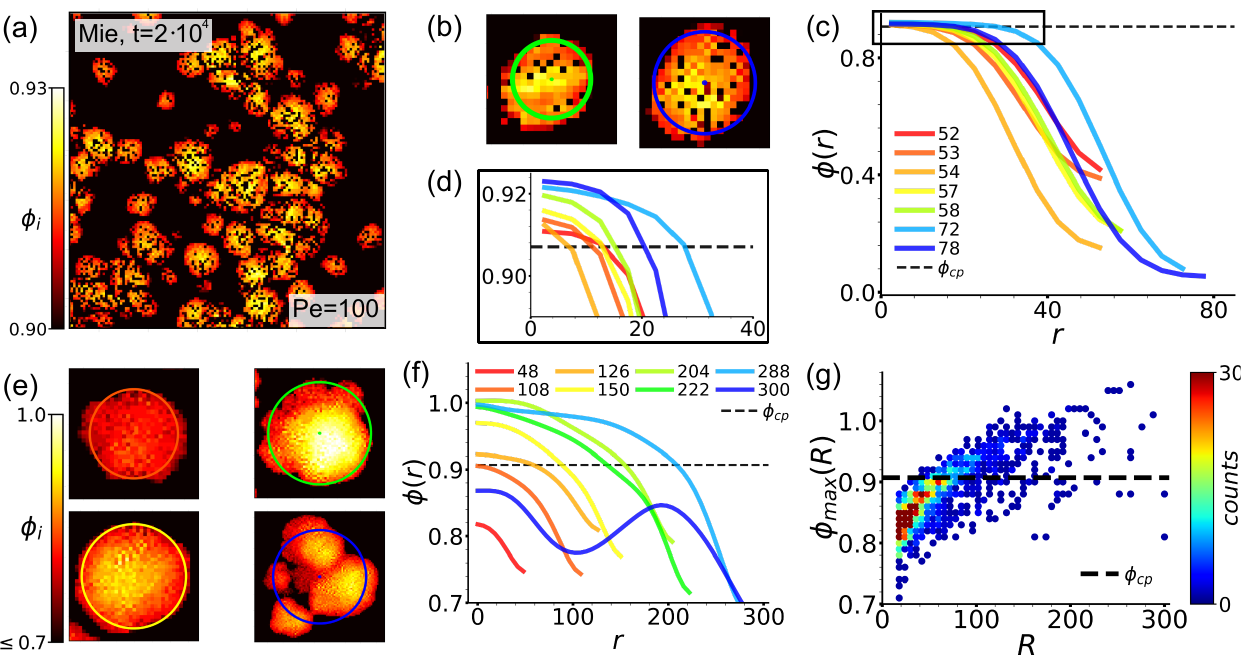}
\caption{{\textbf{Compression and density profiles in isolated clusters.} (\textbf{a}). Local density field $\phi_i$ with $\sigma_{cg}=10\sigma_d$ at $t=2\cdot 10^4$ in an enlarged area of the system in the Mie configuration at ($\rho\sim 0.4$, \mbox{$Pe=100$}). $\phi_i$ is colored according to the left bar. (\textbf{b},\textbf{e}). Local density field in isolated Mie and LJ clusters. In (\textbf{b}), the field colors follow the bar in (\textbf{a}), in (\textbf{e}), its left bar. Dots and circles illustrate centers of mass and assigned radii, and both are colored to match the respective legends in (\textbf{c},\textbf{f}). (\textbf{c},\textbf{f}). Representative density profiles for isolated rounder clusters of increasing dimension in the Mie and LJ configurations, respectively. Curves are colored by cluster radius (see the respective legends). Horizontal lines mark $\phi_{cp}\sim 0.907$. (\textbf{d}). Enlargement of the boxed region in (\textbf{c}). (\textbf{g}). Maximum density $\phi_{max}(R)$ as a function of radius $R$ for isolated rounder LJ clusters sampled every $\tau=10^5$. Each dot represents an $(R, \phi_{max}(R))$ instance that occurred almost once, while its color encodes its frequency over $10$ independent simulation runs (right bar). The horizontal dashed line marks $\phi_{cp}$.}}
\label{fig:fig6}
\end{figure}

As shown in Section~\ref{sec:dens_field}, underlying polarization structures drive bead compression, thus promoting the emergence of peculiar density patterns. In the LJ case, this was evident already in Figure~\ref{fig:fig4} due to the action of a softer repulsive core. The more careful representation of Figure~\ref{fig:fig6}a reveals compressive effects to occur also in the Mie case, although, as expected, in this case they are less intense. In order to better characterize this compressive phenomenology, we analyze density patterns in isolated clusters which, as shown in Section~\ref{sec:pol_prof}, most likely contain inward-pointing topological defects.

Figure~\ref{fig:fig6}b reports two sample clusters in the Mie case colored according to their density fields. Already at this level it is possible to guess a radial-symmetric structure, with $\phi_i$ values decreasing from center (yellow pixels) to boundary (red pixels). A clearer representation of this is provided by Figure~\ref{fig:fig6}c, which reports density profiles $\phi(r)$ for droplets of increasing radius $R$. Here, $r$ denotes the radial distance from the center of mass, which in isolated clusters essentially coincides with the location of defect cores, while radii $R$ are assigned according to where $\phi(r)$ vanishes (dots and circles in Figure~\ref{fig:fig6}b denote centers of mass and assigned radii, and are colored according to the legend in Figure~\ref{fig:fig6}c). These profiles indeed show $\phi(r)$ reducing from the core, where $\phi(0)\sim\phi_{cp}$, to vanishing values at the boundary. Interestingly, the enlargement of the boxed area in Figure~\ref{fig:fig6}c reported in Figure~\ref{fig:fig6}d reveals that at the core generally $\phi(0)>\phi_{cp}$, thus clearly denoting compression is in act, with $\phi(0)$ values slightly increasing with the cluster dimension.

The compressive phenomenology described in the Mie case becomes much more evident in the LJ one. Already at a visual inspection, the selected clusters of increasing dimension reported in Figure~\ref{fig:fig6}e make this manifest (representation is as in Figure~\ref{fig:fig6}b, with reference legend for dots and circles colors now in Figure~\ref{fig:fig6}f). More quantitatively, Figure~\ref{fig:fig6}f reports density profiles for selected clusters of increasing dimension. In the LJ case, $\phi(r)$ decreases much more rapidly at the boundaries, hence the starting value in the panel is $\phi(r)\sim 0.7$. In line with snapshots from Section~\ref{sec:overview} and Appendix~\ref{sec:app}, in the LJ case, it is possible to identify isolated clusters of larger dimensions. Moreover, their decreasing profile character is also now more appreciable closer to the core. The only exception is that of the blue curve referring to one of the largest clusters we sampled, which, up to the largest simulation times we can afford, typically displays a composite structure (see Figure~\ref{fig:fig6}e right bottom). In larger clusters, $\phi(r)>\phi_{cp}$ values survive at further distances from the origin, where, in line with Figure~\ref{fig:fig4}h, a few instances of $\phi(0)\sim 1$ now occur. Note also that, apart from the blue line, the increasing character of $\phi(0)$ with cluster dimension seems to be not only confirmed, but even enhanced.

Support for this last statement is provided by Figure~\ref{fig:fig6}g, where we report the trend of $\phi_{max}(R)\equiv \phi(0)$ as a function of radius $R$ for round LJ clusters. The identification $\phi_{max}(R)\equiv \phi(0)$ is justified by the fact that, as shown by Figure~\ref{fig:fig6}c,f, in isolated clusters, $\phi(r)$ typically peaks at $r\sim0$ for symmetry reasons. Clusters are sampled periodically, waiting enough time for the system to decorrelate. Each dot represents an $(R, \phi_{max}(R))$ instance that occurred almost once, while its color encodes its frequency over $10$ independent simulation runs. While most counts occur for smaller clusters with $\phi_{max}(R)\sim 0.85$, the overall trend is clearly increasing, with the largest clusters generally interested by $\phi_{max}(R)>\phi_{cp}$ values. Unluckily, our current statistics do not allow us to extract a precise functional trend. Still, the overall picture that emerged here is reminiscent of that from \cite{semeraro2025}. There, in fact, we uncovered that domains interested by inward spiral-like or aster-like defects display a decreasing trend similar to the one from Figure~\ref{fig:fig6}f, with $\phi_{max}(R)$ increasing with domain dimension due to strengthened compressive effects (cfr. Figure 4 in \cite{semeraro2025}), which in turn are proven to play a pivotal role in enhanced domain growth.

\section{Entropy Production of Clusters and Aggregates}
\label{sec:ep}

Entropy production of single units is an important subject in active frameworks as it helps to quantify the overall degree of irreversibility of the system. However, parity of active force under time reversal transformation is not uniquely determined  \cite{shankar2018,dabelow2019,fodor2022,byrne2022}. In principle, one could arbitrarily consider the latter either odd or even, as both choices are equally possible and supported by physical interpretations \cite{dabelow2019,fodor2022,byrne2022,oh2023} (for active dumbbells, an odd choice amounts to inverting the identity of head and tail beads in the time-backward evolution). Accordingly, up to negligible boundary terms $\sim\mathcal{O}(1)$, two different bead-wise expressions emerge: 
\begin{equation}
    \dot{{\mathcal S}}^i_+\equiv\lim_{\tau\uparrow\infty}\frac{1}{\tau}\frac{F_a}{k_BT}\int_0^\tau \hat{n}_{i,i+1}(s)\dot{\bm{r}}_i(s)~ds=\lim_{\tau\uparrow\infty}\dot{{s}}^i_+
    \label{eq:s_plus}
\end{equation}
\begin{equation}
    \dot{{\mathcal S}}^i_-\equiv\lim_{\tau\uparrow\infty}\frac{1}{\tau}\frac{F_a}{\gamma k_B T}\sum_{i\neq j}\int_0^\tau \hat{n}_{i,i+1}(s)\nabla_i U(r_{ij}(s))~ds=\lim_{\tau\uparrow\infty}\dot{{s}}^i_-~,
    \label{eq:s_minus}
\end{equation}
where the subscripts $\pm$ denote that the active force is assumed to be even or odd. We remark that $\dot{\mathcal S}^i_+$ is strictly related to the rate of {\it active work}, from which it differs only for a $1/T$ \mbox{factor \cite{dabelow2019, pietzonka2019, keta2021, semeraro2023}}. This is a fundamental observable in active systems as it captures the energy cost to sustain self-propulsion \cite{dabelow2019, mandal2017b, pietzonka2019} and defines the efficiency for active \mbox{engines \cite{pietzonka2019, fodor2021}}. We also remark that the two above expressions are not disconnected as, similarly to \cite{semeraro2024}, one can prove that $\braket{\dot{{\mathcal S}}^i_+}+\braket{\dot{{\mathcal S}}^i_-}=\braket{\dot{ s}^i_+}+\braket{\dot{ s}^i_-}=F_a^2/(\gamma k_B T)$. Therefore, without loss of generality, in the following we focus on $\dot{{s}}^i_+$ for our dumbbells.

\begin{figure}[t!]
\centering
\includegraphics[width=0.8\columnwidth]{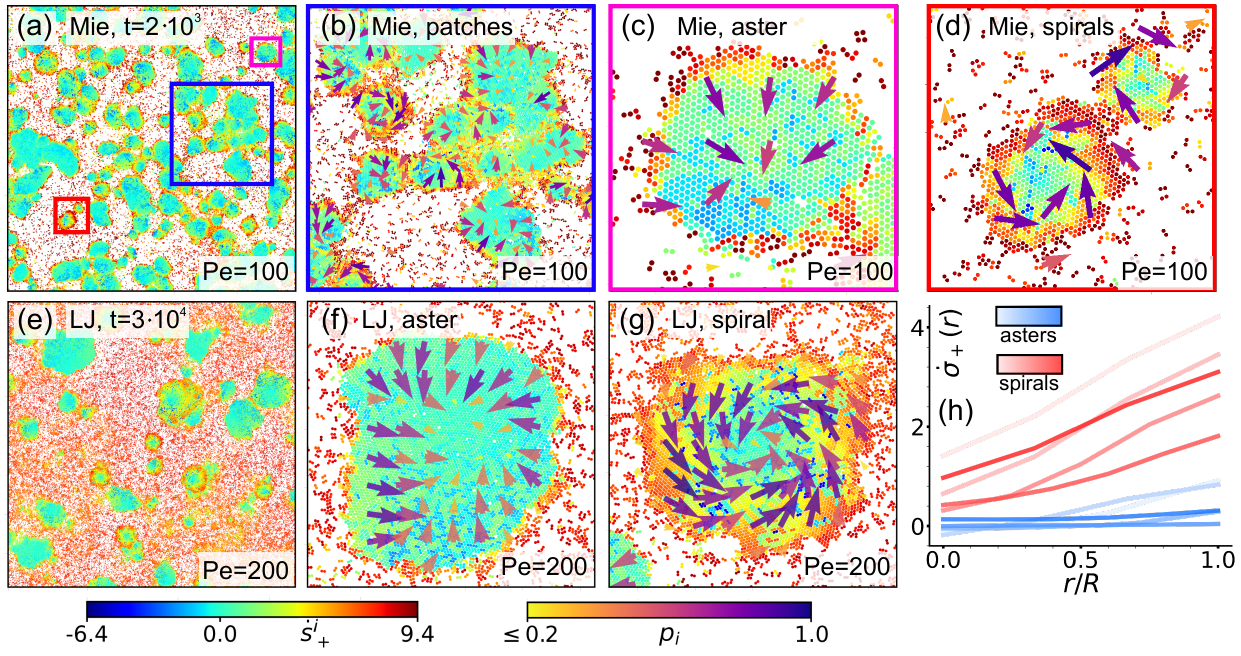}
\caption{{\textbf{Entropy production.} (\textbf{a},\textbf{e}). {Snapshots} of an enlarged Mie at $Pe=100$ and LJ at $Pe=200$ configurations at $\rho\sim 0.4$. Beads are colored according to their $\dot{s}_+^i$ values (left bottom bar). \mbox{(\textbf{b}--\textbf{d}).}~Enlargements of regions delimited in (\textbf{a}) by rectangles with matching colors and, respectively, showing hexatically patched domains, a cluster hosting an aster defect, and two spiraling domains, all with radius $R\sim10$. Beads are colored as in (\textbf{a}). Overlapped, the coarse-grained polarization field $\vec p_i$, colored according to the right bottom bar. (\textbf{f},\textbf{g}). Typical larger clusters in the LJ configuration hosting an aster-like and a spiral-like defect. Representation is as in the top row. In (\textbf{a}--\textbf{g}), $\dot{s}_+^i$ is sampled over $\tau=10^3\tau_{MD}$. (\textbf{h}). Representative entropy profiles $\dot{\sigma}_+(r)$ for rounder clusters of radius $R\sim 20$. Red and blue curves refer to domains hosting an aster or a spiral defect.}}
\label{fig:fig7}
\end{figure}

In Figure~\ref{fig:fig7}a--g we report snapshots of enlarged regions of the system in the Mie and LJ configurations, with beads colored according to their $\dot{s}_+^i$ values. Here, $\dot{s}_+^i$ is sampled over time intervals longer than both inertial $\tau_I=m/\gamma=0.1\tau_{MD}$ and rotational $\tau_a=\gamma\sigma^2_d/(2k_BT)\sim 10^2\tau_{MD}$ timescales. In the following, we consider $\tau=10^3\tau_{MD}$. However, we checked that consistent results are obtained also with the lower $\tau=10^2\tau_{MD}$. Before delving into the comment on such snapshots, we remark that $\dot{s}_+^i$ is sensible to particle aggregation, being on the one side large and positive when dumbbells are driven unhindered by their self-propulsion and on the other large and negative when dragged against their active force. Specular comments apply to $\dot{s}_-^i$, which is large and positive when particles are in close contact and push each other, driven by activity.

With this in mind, Figure~\ref{fig:fig7}a, which focuses on an enlarged Mie configuration, clearly shows that dumbbells in the dilute phase, which can move almost freely in the direction of their self-propulsion for long time intervals, are indeed characterized by $\dot{s}_+^i\gg 0$ values. In line with \cite{chiarantoni2020, martin2021}, we also observe that dumbbells at the boundary of domains, which are characterized by dynamic addition/replacement of units, also display non-vanishing $\dot{s}_+^i$ values. As for dumbbells well within domains, Figure~\ref{fig:fig7}a seems to suggest these to be interested by vanishing values due to the caging effect exerted by close neighbors. However, Figure~\ref{fig:fig7}b--d reveal a more faceted phenomenology. In particular, in Figure~\ref{fig:fig7}b, we report an enlargement of (a) over the blue boxed area, in which composite hexatically patched domains are present. This shows that, while dumbbells well within patches actually display $\dot{s}_+^i\sim 0$ values, those at the boundary between different patches are characterized by positive $\dot{s}_+^i$ values. As hinted in Section~\ref{sec:dens_field}, the interlock between patches is often imprecise, thus generating grain boundaries where dumbbells have a reduced, yet not zeroed, mobility, hence the observed entropy values. In Figure~\ref{fig:fig7}c,d we instead report enlargements of (a) over the magenta and red boxed areas, in which isolated domains of radius $R\sim 10$ hosting an aster-like and spiral-like defect are present. These show an interesting difference between the two defects: in the aster-like cluster, dumbbells at the boundary of (well-inside) the domain are characterized by non-vanishing (vanishing) values. On the contrary, in the spiral-like configuration, dumbbells show $\dot{s}_+^i>0$ values increasing from core to boundary. This reflects how defects affect domain dynamics: aster-like clusters, which lack a tangential component, remain essentially immobile, with push radially compensated, whereas, in spiral-like clusters, the non-vanishing tangential component drives uniform rotations with a translational velocity growing from core to boundary, hence the observed increasing entropy values.

As for the LJ case, Figure~\ref{fig:fig7}e shows that overall comments reported above still apply, apart obviously from the ones concerning grain boundaries between patches, which here are absent. However, as shown by Figure~\ref{fig:fig7}f,g, here the aster-like and spiral-like phenomenology becomes more evident as softer interactions allow the formation of larger isolated clusters hosting a defect. In this regard, in Figure~\ref{fig:fig7}h we report a selection of entropy profiles $\dot{\sigma}_+(r)$ for clusters of radius $R\sim 20$ hosting aster-like (blue curves) or spiral-like (red curves) defects. These are obtained similarly to the density profiles from Section~\ref{sec:dens_prof} and make explicitly evident that, well within, aster-like clusters are indeed characterized by vanishing flat profiles, while spiral-like by positive increasing ones. In addition, the latter start from non-vanishing values already at the core and consistently keep larger values than aster-like ones, even at the boundary. Overall, these observations suggest a pathway alternative to dynamical measurements, as the ensthropy from \cite{petrelli2018}, to distinguish different types of defects based on irreversibility criteria.

\section{Conclusions}
\label{sec:conclusions}

In this paper, we studied two-dimensional tail--head polar active dumbbell systems undergoing MIPS to investigate the interplay between domain morphology, shape and growth, polarization patterns, compression, and irreversibility. Investigation was performed through numerical simulations and coarse-graining procedures in two settings obtained by varying tail--head rigidity and interaction rule: a rigid one, with fixed tail--head distance and a hard Mie repulsive pairwise interaction; a softer one, in which tail--head distance can slightly oscillate and beads interact through a softer repulsive Lennard--Jones potential.

Our analysis uncovered distinct differences, all ultimately stemming from contrasting mechanisms in action: Mie interactions favor rigid dumbbells interlocking, while LJ ones promote bead sliding and compression. As a consequence, and in contrast to the well-characterized hard-interacting picture, our results revealed that softer interactions give rise to blurred hexatic patterns, polarization patterns extended across entire hexatically varied domains and strong compression effects. These were in turn ascribed to the action of boundary dumbbells exerting an inward pressure. Analysis of the internal structure of isolated domains revealed the consistent presence of inward-pointing topological defects that emerge naturally as a consequence of the initial nucleation mechanism. These drive cluster compression and generate non-trivial density profiles, whose amplitude and extension are enhanced in the softer setting. Investigation of entropy production additionally showed that dumbbells close to grain boundaries between hexatic patches display larger entropy values due to their reduced, yet not zeroed, mobility in these regions. Moreover, clusters with an aster-like (spiral-like) defect are found to be characterized by a flat (increasing) entropy profile. This at the same time mirrors how defects affect cluster dynamics and suggests an alternative pathway based on irreversibility to distinguish topological defects.

Overall, our study sheds light on the effect of interaction strength and polarization–compression interplay on evolution in polar particle-based active models. Moreover, it also offers the opportunity to establish connections with our recent study of a continuum polar active field model from \cite{semeraro2025}, thus providing preliminary support to the idea that the top-down description proposed there can indeed serve as a valid picture of phase separation in polar particle systems. However, our efforts in this direction must be interpreted as just a first qualitative step towards linking active dumbbell systems, or more general particle systems displaying polar features, with active field models. A more quantitative analysis, which we leave for future implementation, is in fact of order. Ultimate connection will be provided by a rigorous coarse-graining procedure of the particle-based model from Section~\ref{sec:model} into a continuum field theory, which, to our knowledge, is still lacking.

The disappearance of grain boundaries in clusters with softer interactions raises further questions regarding the internal structure of the ordered phases of these systems, that would be interesting to address in the future. In fact, similarly to disks, dumbbells arrange in hexatic structures hosting topological defects, such as dislocations and disclinations, which are of relevant interest in the study of the passive melting \cite{olson2002,gerbode2008, gerbode2010, gerbode2010b} and jamming \mbox{transitions \cite{reichhardt2014}} in 2D and, more generally, in the mechanical response to external \mbox{shearing \cite{olson2002, carenza2025}}. Therefore, it would be interesting to extend the investigation started in this paper to see how the overall passive scenario is affected by activity, interaction rule, and particle shape/elongation.

\appendix
\section[\appendixname~\thesection]{Supplementary Figures}
\label{sec:app}

\begin{figure}[H]
\centering
\includegraphics[width=0.4\columnwidth]{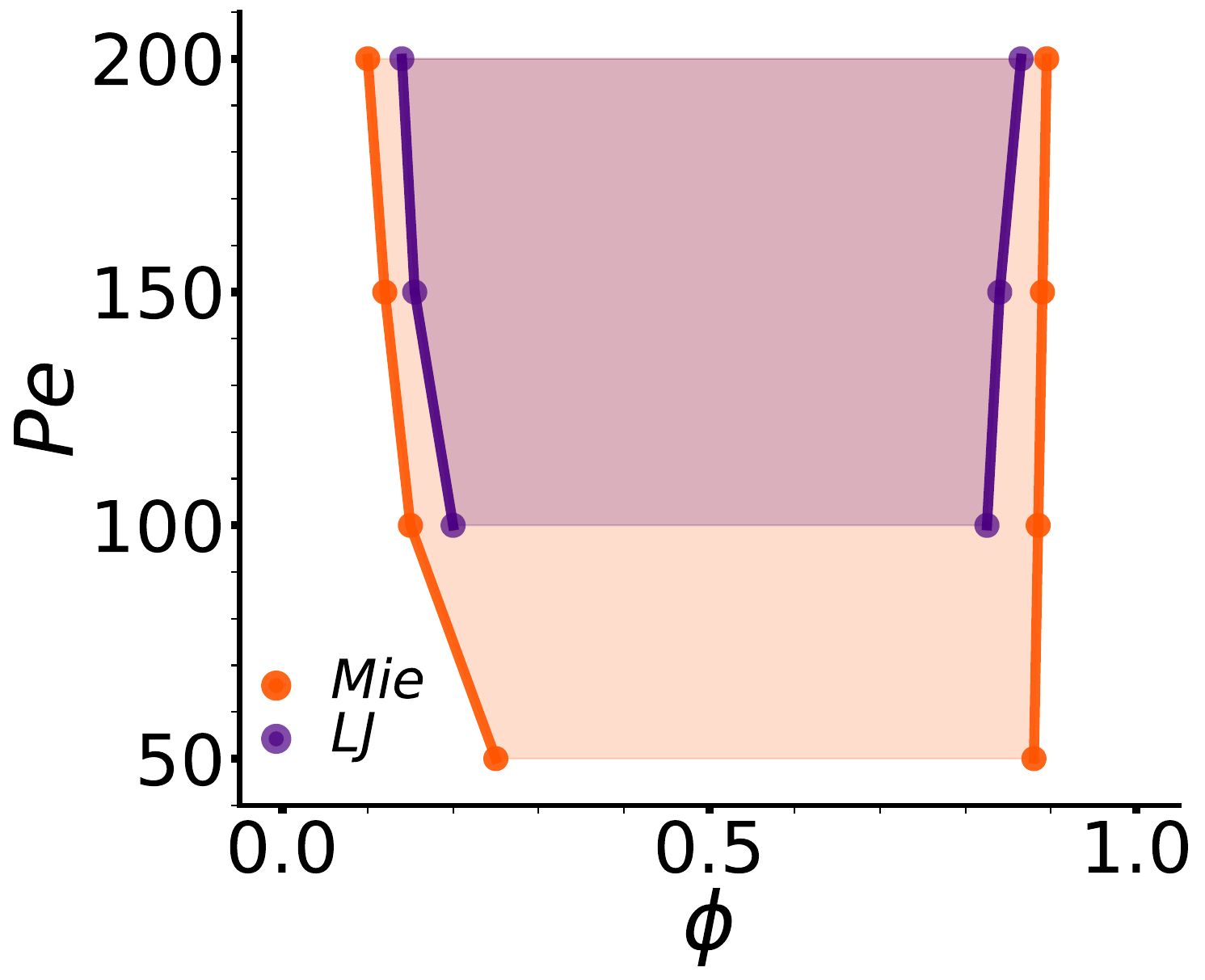}
\caption{{{\textbf{Phase diagrams.} Indicative phase diagram in the $(\phi, Pe\geq 50)$ plane in the Mie (orange) and LJ (violet) configuration. The colored areas denote regions where phase separation occurs. The complete phase diagram for the Mie case is studied in detail in \cite{cugliandolo2017, petrelli2018}. In the LJ case, no phase separation is observed for $Pe<100$ at the explored densities $\phi\sim$ 0.4--0.5.)}}}
\label{fig:figs1}
\end{figure}

\begin{figure}[H]
\centering
\includegraphics[width=0.8\columnwidth]{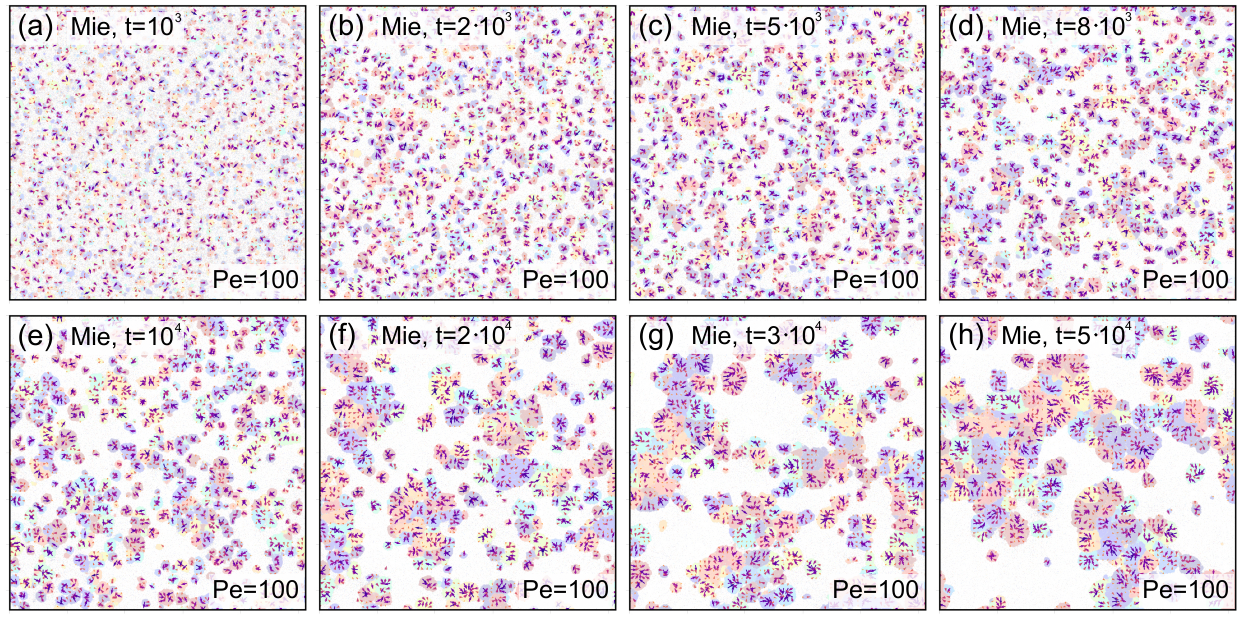}
\caption{{\textbf{The local polarization density field in the Mie configuration.} \mbox{(\textbf{a}--\textbf{h}).} Local polarization density field $\vec p_i$ with $\sigma_{cg}=10\sigma_d$ over the entire system at ($\rho\sim 0.4$,~$Pe=100$) in the Mie configuration at increasing times.  The overall representation is as in Figure~\ref{fig:fig3}: the field is represented as arrows colored according to their magnitude $p_i=|\vec p_i|$, arrows such that $ p_i<0.2$ are removed, backgrounds report corresponding hexatic snapshots.}}
\label{fig:figs2}
\end{figure}

\begin{figure}[H]
\centering
\includegraphics[width=0.8\columnwidth]{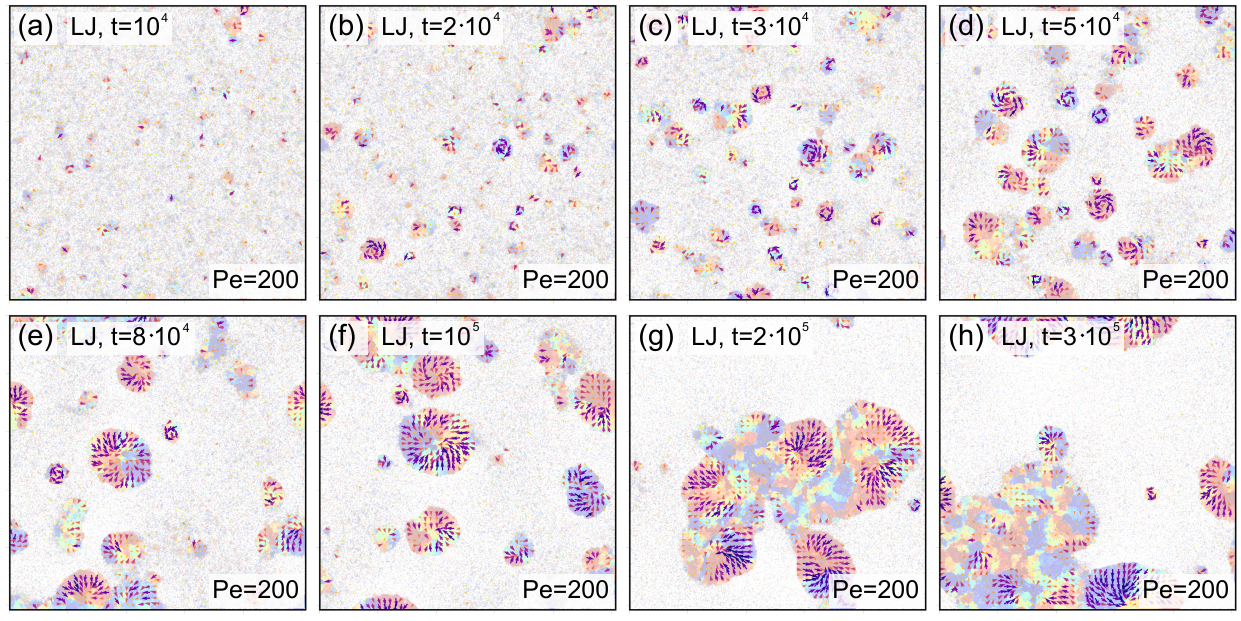}
\caption{{\textbf{The local polarization density field in the LJ configuration.} \mbox{(\textbf{a}--\textbf{h}).} Local  polarization density field $\vec p_i$ with $\sigma_{cg}=10\sigma_d$ over the entire system at ($\rho\sim 0.4$,~$Pe=200$) in the LJ configuration at increasing times. Representation is as in Figure~\ref{fig:figs2}.}}
\label{fig:figs3}
\end{figure}

In this appendix we report supplementary figures
supporting our main discussion. In particular, combining available data, in Figure~\ref{fig:figs1} we report an indicative phase diagram in the $(\phi,Pe\geq50)$ plane for both Mie and LJ configurations. In Figures~\ref{fig:figs2} and~\ref{fig:figs3} we instead report snapshots of the polarization density fields at different times in the Mie and LJ configurations, respectively.

\section*{Acknowledgments}

Numerical calculations have been made possible through a Cineca--INFN agreement, providing access to HPC resources at CINECA. All authors acknowledge support from the INFN/FIELDTURB467 project,  from MUR projects PRIN 2020/PFCXPE, PRIN 2022/HNW5YL, and PRIN 2022 PNRR/P20222B5P9, and Quantum Sensing and Modelling for One-Health (QuaSiModO).

\section*{Bibliography}

\bibliography{Bibliography_arxiv.bib}

\end{document}